# Temperature Measurement Based on Electron Spin Resonance of Magnetic Nanoparticles

Shuai Wang (王帅)[1,2,3], Jing Zhong (钟景)[4,‡], and Wenzhong Liu (刘文中)[1,2,3,†]

[1] *School of Artificial Intelligence and Automation, Huazhong University of Science and Technology, Wuhan 430074, People's Republic of China*
[2] *Belt and Road Joint Laboratory on Measurement and Control Technology, Huazhong University of Science and Technology, Wuhan 430074, People's Republic of China*
[3] *Key Laboratory of Image Processing and Intelligent Control, Huazhong University of Science and Technology, Ministry of Education, Wuhan 430074, People's Republic of China*
[4] *School of Instrumentation and Optoelectronic Engineering, Beihang University, Beijing 100191, People's Republic of China*

Magnetic nanoparticles (MNPs) have excellent magnetic-temperature characteristic. However, current temperature measurement based on MNPs is interfered by concentration. Utilizing the electron spin resonance (ESR), we propose a highly sensitive temperature measurement method without concentration coupling. The anisotropic field is affected by temperature, thus affecting the *g*-value. The influence of the MNP concentration, size, and the data analysis method on temperature estimation are studied. The optimal temperature sensitivity is achieved with 15-nm MNPs while Gaussian smoothing method allows an optimal accuracy at Fe concentration of 5 mg/ml with a root mean squared error of 0.07 K.

Temperature is an important characterization of the physiological activities of cells [1-4]. For example, DNA replication, transcription and RNA processing, as well as its structural separation by the nuclear membrane [5], may accelerate the average temperature of nucleus by 0.96 K compared to the cytoplasm [6]. A novel thermometer with extremely high sensitivity is critical to resolve the sub-degree temperature change caused by cell metabolism at a single cell level [7].

Magnetic means are promising approaches for in vivo temperature measurement [8], due to the magnetic transparence of human body. However, the current magnetic temperature measurement methods used in micro/nanoscale applications are affected by the concentration of the temperature probe [9,10]. The complex biological structure of cells and the cell metabolism [11-13] lead to the instability of temperature probes concentration, so it is difficult to achieve high-precision temperature measurement of organelles, including mitochondria [14,15]. Therefore, a temperature measurement method independent of probe concentration is required to promote the research into cell bio thermophysics.

Recently, magnetic nanoparticles (MNPs) have been used as temperature sensors for thermometry, showing great promises in remote and in vivo temperature measurement. [8,16,17]. The Weaver and Liu research groups have comprehensively investigated temperature measurement methods based on MNP magnetization [18-22]. Different magnetic resonance imaging (MRI) contrast agents, such as iron oxide nanoparticles, gadolinium and doped ferrite nanoparticles, have been used as thermometers for temperature measurement [23-26]. However, both the magnetization and the MRI parameters are severely affected by the particle concentration so that the high-precision temperature measurement remains challenging.

In this letter, we propose a new approach for remote and sensitive temperature measurement based on the electron spin resonance (ESR) spectrum of MNPs (Fe$_3$O$_4$) [27-29]. The magnetic anisotropy of MNPs decreases as the temperature increases, resulting in a decrease in the magnetic anisotropy field, thus increasing the resonance field [30-35]. The *g*-value is one-to-one correspondence with the resonance field. By studying the changes in the *g*-value with temperature, a temperature measurement model can be established, allowing temperature information to be obtained. Experiments have found that the MNP concentration has no effect on *g*-value within the observed temperature range, thus eliminating the interference of concentration on the temperature sensitivity of the probe. According to this experimental phenomenon, we optimize temperature measurement parameters, including the particle concentration and size, and compare different data analysis algorithms. Finally, we realize a highly sensitive ESR temperature measurement method.

The physical model of ESR-spectrum-based temperature measurement depends on the influence of the superposition of the internal field and the external field of the particle. In a single-domain nanoparticle, the magnetic anisotropy energy may easily become comparable with the thermal energy, and the thermal fluctuation is particularly important [30,31,33,34]. The MNPs in ESR are affected by thermal fluctuations and the changes in the anisotropic axis orientation distribution caused by the particle orientation mobility under the action of a magnetic field [36,37]. As the temperature increases, the thermal fluctuation of the magnetic moment increases whereas the contribution of the anisotropy field to the resonance magnetic field decreases. The main axis of an ellipsoidal particle coincides with one of the local magnetic axes. It has been demonstrated that in this case, the resonance field can be expressed as [38-41]

$$H_r = H_0 + H_a + H_d, \qquad (1)$$

where $H_0$ is the applied magnetic field, and $H_a$, $H_d$ are, respectively, the projection of magnetocrystalline anisotropy field and demagnetizing field in the direction of $H_0$.

To explore the influence of temperature on the resonance field, Eq. (1) is further studied, and the applied magnetic field $H_0$ can be expressed as

$$H_0 = \frac{h\upsilon}{g_e \beta}, \quad (2)$$

where $h = 6.626\times10^{-34}$ J/s is the Planck's constant, $\upsilon$ is the microwave frequency, $g_e = 2.00232$ is the g-value of free electrons, and $\beta = 9.274\times10^{-24}$ J/T is the Bohr magneton. As the iron oxide nanocrystals have a cubic symmetric structure, the easy directions of magnetization correspond to [1 1 0] axes [36,42], and the magnetocrystalline anisotropy field $H_a$ can be expressed as

$$H_a = \frac{2K}{M_s}\frac{L_4(\xi)}{L_1(\xi)}\left(\frac{5}{4}\sin^2 2\vartheta + \frac{5}{4}\sin^4\vartheta\sin^2 2\varphi - 1\right), \quad (3)$$

with

$$\begin{cases} L_4(\xi) = 1 + \frac{35}{\xi^2} - \left(\frac{10}{\xi} + \frac{105}{\xi^3}\right)L_1(\xi) \\ L_1(\xi) = \coth(\xi) - \frac{1}{\xi} \\ \xi = \frac{M_s V H_r}{k_B T} \end{cases},$$

where $K$ is the anisotropy constant, $M_s$ is the saturation magnetization, $\vartheta$ is the angle between the anisotropy axis and the external magnetic field, $\varphi$ is the azimuth angle, $V = \pi D^3/6$ is the particle volume, $D$ is the core particle size, $k_B = 1.38\times10^{-23}$ J/K is the Boltzmann constant, and $T$ is the particle temperature. The demagnetizing field $H_d$ can be expressed as

$$H_d = \frac{\mu_0 M_s}{2}(N_\parallel - N_\perp)(3\cos^2\vartheta - 1)L_1(\xi), \quad (4)$$

where $\mu_0 = 4\pi\times10^{-7}$ T·m/A is the vacuum permeability, and $N_\parallel$, $N_\perp$ are demagnetization factors for the directions parallel and perpendicular, respectively, to the major axis of the ellipsoid. The g-value can be obtained by

$$g = \frac{h\upsilon}{\beta H_r}. \quad (5)$$

Eqs. (1)–(5) indicate that temperature mainly affects the magnetic anisotropy field, consequently changing the resonance field and g-value. For MNPs, within the physiological temperature range, the g-value changes monotonously with the probe temperature. Based on this, the model of temperature dependent g-value can be constructed for temperature determination. The interference of MNP concentration is excluded from this model, thus allowing the remote temperature measurement with high-precision.

Based on this model, we experimentally investigated the influence of the particle parameters, including core diameter and concentration, on the temperature dependent g-value. The particles used in the experiment are SHP-5, SHP-10, SHP-15, and SHP-20, with core diameter of 5, 10, 15, and 20 nm, respectively. They are water-suspended iron oxide nanocrystals with amphiphilic polymer coating (Ocean Nanotech, USA). Their surfaces have the same negative charge, which will hinder the formation of aggregates through the strong net particle repulsive force. The ESR spectrum is scanned using JES-FA200 (JEOL, Japan), which works in X-band and enables accurate temperature control.

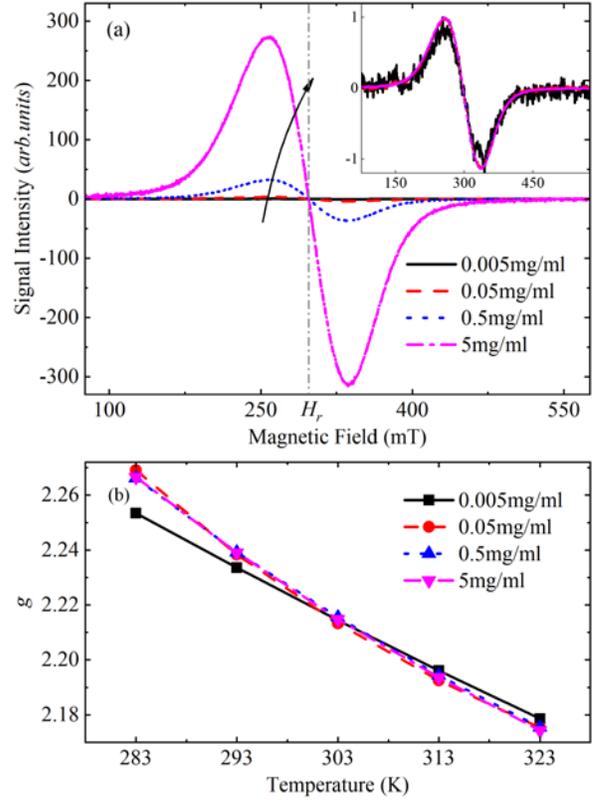

FIG. 1. Concentration dependent ESR spectra and g-value. (a) ESR spectra of different concentrations of SHP-15. The arrow points in the direction of increasing concentration. The inset shows the normalized spectrum. (b) Temperature-dependent g-value for different concentrations of SHP-15.

To investigate the influence of the MNP concentration on the g-value, the ESR spectra of SHP-15 MNPs with Fe concentrations of 0.005, 0.05, 0.5, and 5 mg/ml are measured at 313 K. Experimental results are shown in Fig. 1(a). The resonance magnetic field $H_r$ is the field corresponding to the center zero point of the ESR spectrum, as shown by the gray dotted line. The difference between the peak and the trough in the ESR spectra (Peak to Trough) has been extensively used to quantify the signal intensity [43-45], which is significantly affected by the particle concentration. As the concentration increases, the intensity of the spectrum increases correspondingly. However, lower concentrations exhibit higher fluctuations in their spectra, which severely disturbs the extraction of g-value. The normalized spectra of the four curves are shown in the inset of Fig. 1(a), which indicates that the resonance field does not change with concentration.

Figure 1(b) shows the temperature-dependent g-value for different-concentration SHP-15 MNPs. When the concentration exceeds 0.05 mg/ml, the three curves of g-value versus temperature collapse to a single curve. It indicates that the particle concentration does not affect the g-value. The explanation is that the g-value is an intrinsic property of matter, which can only be affected by the interactions between free electrons and other electron spins or nuclear spins around the local field. However, the curve for the 0.005 mg/ml sample exhibits a large deviation compared with the high-concentration sample. It might be caused by the system noise at low concentrations, leading to a large interference in extracting *g*-value. Therefore, to improve the temperature measurement accuracy, the sample concentration should be comparably high to have a nice signal-to-noise ratio.

The physical model indicates that the MNP size has significant influence on the *g*-value of the spectrum. Four different particles of SHP-5, SHP-10, SHP-15, and SHP-20 with the same Fe concentration of 0.5 mg/ml are compared to observe their temperature sensitivity of ESR spectra. Figure 2(a) shows the transmission electron microscope (TEM) images of the four different particles, exhibiting the uniform particle size. The spectra of the four different MNPs scanned at room temperature are presented in Fig. 2(b). The spectrum of SHP-5 has the strongest intensity, with a very narrow spectral line and a *g*-value close to $g_e$. As the particle size increases, the intensity gradually decreases, with the spectral line broadening and the *g*-value deviating from $g_e$. This is because the internal field of small particles is dominated by the isotropic property. Herein, electrons can be regarded as free electrons, so the spectral line is very narrow and approximately Lorentzian [30]. As the particle size increases, the anisotropic property begins to dominate [33,40], the spectral line broadens, and the *g*-value shifts. In addition, the large-particle-size sample contains fewer particles under the same Fe concentration. Thus, the spin concentration is correspondingly lower, leading to the weaker spectrum intensity.

The experimental data of the *g*-value of the four different MNPs versus temperature are shown in Fig. 2(c). Within the temperature range of 283–323 K, SHP-5 exhibits minor change in *g*-value, representing a bad temperature sensitivity. On contrast, the data of the other MNPs show considerable changes in *g*-values of 0.0743, 0.09179 and 0.07658 for 10-nm, 15-nm and 20-nm MNPs, respectively. The greatest change occurs in the 15-nm particles, representing the best temperature sensitivity. The *g*-values of 10-nm and 15-nm MNPs are relatively similar. The reason may be that the two particles have the similar particle size, and their size distributions are almost the same.

After analyzing the influence of particle concentration and size on the temperature sensitivity of the *g*-value, the result of ESR spectral temperature measurement is evaluated through temperature variation experiment. According to the results of Fig. 2(c), SHP-10 and SHP-15 MNPs with the same Fe concentration of 5 mg/mL are specifically selected as the temperature-sensitive probes for comparison. Their spectra are scanned in the physiological temperature range of 283–323 K. The original spectra are shown in Figs. 3(a) and 3(c). As the temperature increases, the spectral lines shift to a higher field. The inset shows that there is still some fluctuation, reflecting the Gaussian noise introduced by the system.

In order to reduce the noise interference, different optimization algorithms are used to fit the spectra. The fitted results are shown in Figs. 3(b) and 3(d). The red-solid lines are the results of Gaussian smoothing of the original spectra. These lines show that the noise is greatly reduced, and the spectra tend to be smooth. In addition, the original spectral

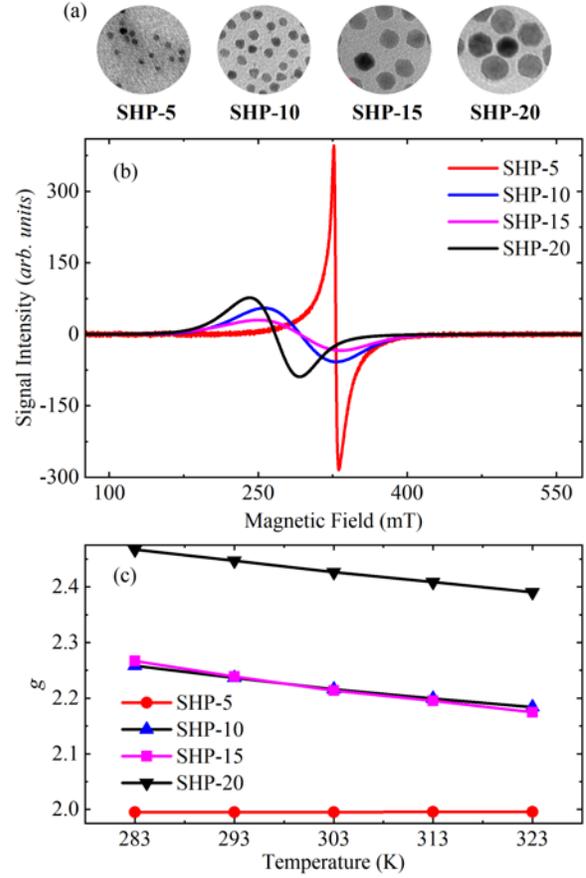

FIG. 2. Particle-size dependence of the ESR spectra and *g*-value. (a) TEM images, (b) spectra at room temperature, and (c) temperature sensitivity of the *g*-value for the four particles.

line has a Gaussian line shape. So the Gaussian model is used to fit the original spectral line with the Levenberg–Marquardt algorithm (L-M fitting), which can be written as

$$f(H) = \left( A \cdot \exp\left( -\frac{(H-H_r)^2}{C^2} \right) \right)'$$
$$= A \cdot \exp\left( -\frac{(H-H_r)^2}{C^2} \right) \cdot \left( -2\frac{H-H_r}{C^2} \right), \quad (6)$$

where $A$, $H_r$, and $C$ are constants. The fitted results are shown as blue-dot-dash lines. The insets of Figs. 3(b) and Fig. 3(d) show that the L-M fitting curves produce a certain deviation from the original spectra. However, temperature inversion focuses on the relative change in the *g*-value, so these deviations can be ignored. The *g*-values extracted by the different methods are shown in Figs. 3(e1–e3) and 3(f1–f3) (red solid dots). An iterative optimization algorithm is used to fit the *g*-value, and the fitting equation can be simplified by the physical model as follows

$$g_{fit} = \frac{a}{T^2} + \frac{b}{T} + c, \quad (7)$$

where *a*, *b*, and *c* are constants. The black dot-dash lines in Figs. 3(e1–3) and 3(f1–3) are the fitted results, which fit very well with the calculated *g*-values. The temperature measurement model can be obtained by solving the fitting

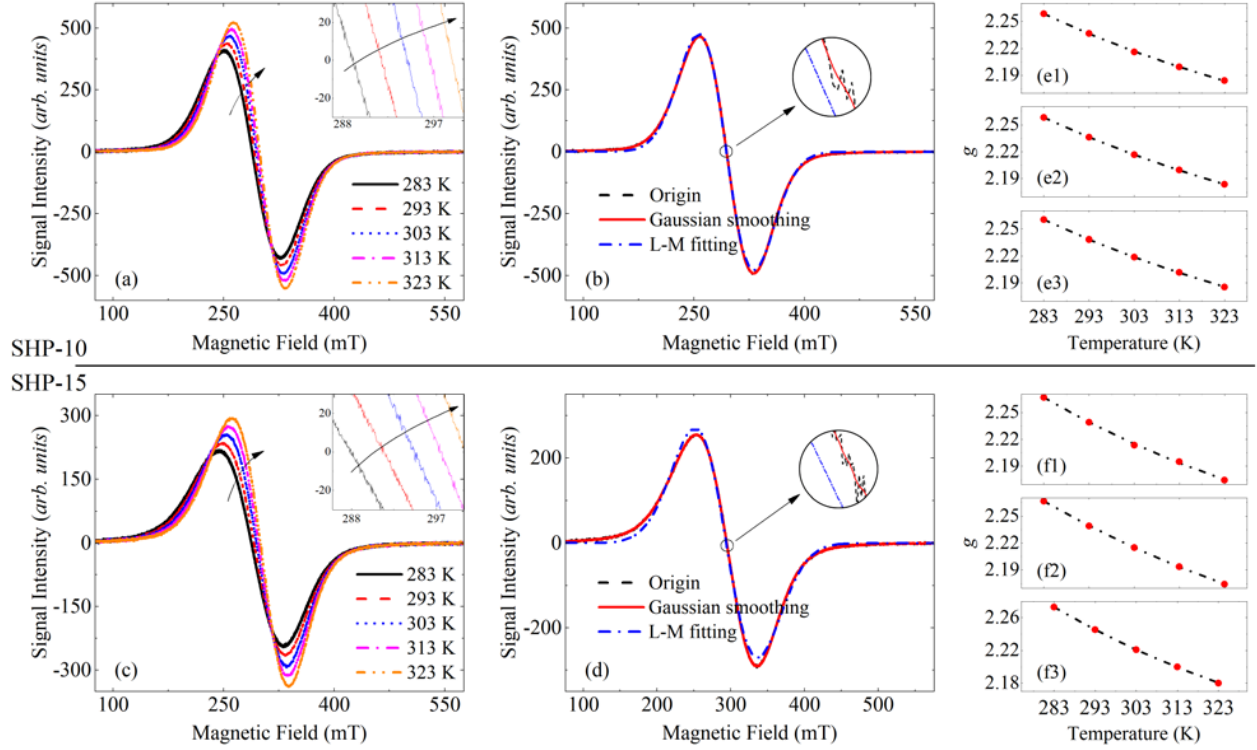

FIG. 3. Variable temperature test results of SHP-10 and SHP-15. (a) and (c) are the original ESR spectra of the two samples. The arrow points in the direction of temperature rise. (b) and (d) are the fitting spectra at a temperature of 303 K. The black-dashed line is the original spectrum. The red-solid line is the Gaussian smoothing spectrum, and the blue-dot-dash line is the L-M fitting spectrum (the model has a Gaussian line shape). (e1-e3) and (f1-f3) are the temperature dependence of $g$-value of the two samples with different methods. Parts (1), (2), (3) are the $g$-values extracted from the original spectrum, Gaussian smoothing spectrum, and L-M fitting spectrum, respectively. Red dots denote experimental values and the black-dot-dash line is the optimal fitting curve.

formula as follows

$$T^* = \frac{1}{\sqrt{\frac{b^2}{4a} + \frac{g-c}{a}} - \frac{b}{2a}} \ . \quad (8)$$

The temperature errors obtained by Eq. (8) are shown in Fig. 4. In overall, the temperature measurement accuracy of SHP-15, shown in Fig. 4(b), is better than that of SHP-10, shown in Fig. 4(a). Comparing the temperature measurement errors of the three methods, the original spectrum gives the largest errors whereas Gaussian smoothing spectrum gives the smallest errors. L-M fitting spectrum produces slightly larger errors than Gaussian smoothing spectrum. Figure 4(c) shows the root mean squared error (RMSE) of the temperature error of the two particles obtained by the different methods. The minimum RMSEs of SHP-10 and SHP-15 reach 0.12 K and 0.07 K, respectively.

In this work, the MNP paramagnetic markers are used as temperature probes, and finally achieve the high-precision ESR temperature measurements with an optimal RMSE of 0.07 K successfully. Considering that the magnetic anisotropy of MNPs is affected by the temperature, a theoretical model of the temperature dependent resonance field was deduced. Our findings show that the particle concentration has no effects on the temperature sensitivity of the $g$-value, which is a great advantage of ESR-based temperature measurement. The particle size is one of the key parameters affecting the temperature sensitivity of the $g$-value. Experiments have found that, among particle sizes ranging from 5–20 nm, the 15-nm MNPs have the best temperature sensitivity. And SHP-10 and SHP-15 MNPs have the similar $g$-value. As the ESR signal intensity is directly related to the particle concentration [45,46], we used high-concentration SHP-10 and SHP-15 samples in temperature dependent experiments. Due to the fluctuation caused by the system noise, different data analysis methods are used, including Gaussian smoothing method and L-M fitting method. By comparing the $g$-value extraction effects of these method, we find that the optimal temperature measurement result is obtained by using the Gaussian smoothing method, and the RMSE of temperature reaches 0.07 K. Furthermore, this method has a simplified magnetics-based temperature measurement model, which is very important in further application. Unfortunately, due to the limitations of the current ESR equipment, the experiments are only carried out using an X-band spectrometer. In principle, it is also of great interest to investigate the temperature measurement accuracy in higher or lower band.

In addition, ESR-based measurement, one of the most sensitive methods for concentration quantification at present, is frequently used as a powerful tool to support the positioning of MRI contrast agents [47]. Therefore, this method provides an interesting approach for in-vivo temperature imaging with promising application prospects in the field of biomedicine.

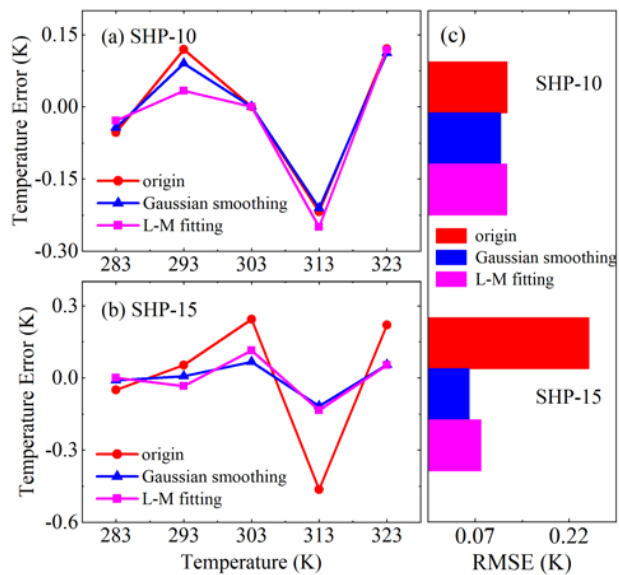

FIG. 4. Temperature error of $g$-value inversion. Error of (a) SHP-10 and (b) SHP-15 particles obtained by original spectrum (red line), Gaussian smoothing spectrum (blue line), and L-M fitting spectrum (magenta line). (c) RMSE of the temperature.

The authors are grateful to the National Center for Magnetic Resonance in Wuhan. This work was supported by the National Natural Science Foundation of China (Grant No. 61973132), the Key Project of Hubei Province (Grant No. 2020BHB020), and the Interdisciplinary Program of Wuhan National High Magnetic Field Center (Grant No. WHMFC202103), Huazhong University of Science and Technology.

‡ zhongjing@buaa.edu.cn
† lwz7410@hust.edu.cn

[1] R. S. Seymour, Biosci. Rep. **21**, 223 (2001).
[2] A. Bahat, I. Tur-Kaspa, A. Gakamsky, L. C. Giojalas, H. Breitbart, and M. Eisenbach, Nat. Med. **9**, 149 (2003).
[3] D. A. Warner and R. Shine, Nature **451**, 566 (2008).
[4] D. Patel and K. A. Franklin, Plant Signaling & Behavior **4**, 577 (2009).
[5] A. I. Lamond and W. C. Earnshaw, Science **280**, 547 (1998).
[6] K. Okabe, N. Inada, C. Gota, Y. Harada, T. Funatsu, and S. Uchiyama, Nat Commun **3**, 705 (2012).
[7] H. Zhou, M. Sharma, O. Berezin, D. Zuckerman, and M. Y. Berezin, Chemphyschem **17**, 27 (2016).
[8] K. Wu, D. Su, R. Saha, J. Liu, V. K. Chugh, and J.-P. Wang, ACS Appl. Nano Mater. **3**, 4972 (2020).
[9] N. W. Lutz and M. Bernard, iScience **23**, 101561 (2020).
[10] H. F. Rodrigues, G. Capistrano, and A. F. Bakuzis, Int. J. Hyperthermia **37**, 76 (2020).
[11] M. G. Vander Heiden, L. C. Cantley, and C. B. Thompson, Science **324**, 1029 (2009).
[12] W. Bailis, J. A. Shyer, J. Zhao, J. C. G. Canaveras, F. J. Al Khazal, R. Qu, H. R. Steach, P. Bielecki, O. Khan, R. Jackson *et al.*, Nature **571**, 403 (2019).
[13] J. M. Yang, H. Yang, and L. Lin, ACS Nano **5**, 5067 (2011).
[14] R. Pinol, C. D. Brites, R. Bustamante, A. Martinez, N. J. Silva, J. L. Murillo, R. Cases, J. Carrey, C. Estepa, C. Sosa *et al.*, ACS Nano **9**, 3134 (2015).
[15] Z. Wang, X. Ma, S. Zong, Y. Wang, H. Chen, and Y. Cui, Talanta **131**, 259 (2015).
[16] A. Akbarzadeh, M. Samiei, and S. Davaran, Nanoscale Res Lett **7**, 144 (2012).
[17] P. Tartaj, M. D. Morales, S. Veintemillas-Verdaguer, T. Gonzalez-Carreno, and C. J. Serna, J. Phys. D: Appl. Phys. **36**, R182 (2003).
[18] J. Zhong, J. Dieckhoff, M. Schilling, and F. Ludwig, J. Appl. Phys. **120**, 1559 (2016).
[19] J. Zhong, W. Liu, Z. Du, P. César de Morais, Q. Xiang, and Q. Xie, Nanotechnology **23**, 075703 (2012).
[20] J. Zhong, W. Liu, L. Kong, and P. C. Morais, Sci. Rep. **4**, 6338 (2014).
[21] A. M. Rauwerdink, E. W. Hansen, and J. B. Weaver, Phys. Med. Biol. **54**, L51 (2009).
[22] J. B. Weaver, A. M. Rauwerdink, and E. W. Hansen, Med. Phys. **36**, 1822 (2009).
[23] J. H. Hankiewicz, Z. Celinski, K. F. Stupic, N. R. Anderson, and R. E. Camley, Nat Commun **7**, 12415 (2016).
[24] N. A. Alghamdi, J. H. Hankiewicz, N. R. Anderson, K. F. Stupic, and Z. Celinski, Phys. Rev. Appl. **9**, 054030 (2018).
[25] J. H. Hankiewicz, J. A. Stoll, J. Stroud, J. Davidson, K. L. Livesey, K. Tv Rdy, A. Roshko, S. E. Russek, K. Stupic, and P. Bilski, J. Magn. Magn. Mater. **469**, 550 (2018).
[26] Y. Zhang, S. Guo, P. Zhang, J. Zhong, and W. Liu, Nanotechnology **31**, 345101 (2020).
[27] V. K. Sharma and F. Waldner, J. Appl. Phys. **48**, 4298 (1977).
[28] C. T. Hseih, W. L. Huang, and J. T. Lue, J. Phys. Chem. Solids **63**, 733 (2002).
[29] R. Krzyminiewski, T. Kubiak, B. Dobosz, G. Schroeder, and J. Kurczewska, Curr. Appl. Phys. **14**, 798 (2014).
[30] R. Berger, J. C. Bissey, and J. Kliava, J. Phys.: Condens. Matter **12**, 9347 (2000).
[31] Y. L. Raikher and V. I. Stepanov, J. Magn. Magn. Mater. **149**, 34 (1995).
[32] F. Gazeau, V. Shilov, J. C. Bacri, E. Dubois, F. Gendron, R. Perzynski, Y. L. Raikher, and V. I. Stepanov, J. Magn. Magn. Mater. **202**, 535 (1999).
[33] Y. L. Raikher and V. V. Stepanov, Phys. Rev. B **50**, 6250 (1994).
[34] Y. L. Raikher and V. I. Stepanov, J. Magn. Magn. Mater. **316**, 417 (2007).
[35] B. R. McGarvey, J. Phys. Chem. **61**, 1232 (1957).
[36] F. Gazeau, J. C. Bacri, F. Gendron, R. Perzynski, Y. L. Raikher, V. I. Stepanov, and E. Dubois, J. Magn. Magn. Mater. **186**, 175 (1998).
[37] R. H. Kodama, A. E. Berkowitz, E. J. McNiff, Jr., and S. Foner, Phys. Rev. Lett. **77**, 394 (1996).
[38] R. S. De Biasi and T. C. Devezas, J. Appl. Phys. **49**, 2466 (1978).
[39] E. Schlömann, J. Phys. Chem. Solids **6**, 257 (1958).
[40] R. Berger, J. Kliava, J. C. Bissey, and V. Baietto, J. Phys.: Condens. Matter **10**, 8559 (1998).
[41] R. Berger, J. Kliava, J. C. Bissey, and V. Baietto, J. Appl. Phys. **87**, 7389 (2000).
[42] J. B. Birks, Proc. Phys. Soc. London, Sect. B **63**, 65 (1950).
[43] S. I. Dikalov, I. A. Kirilyuk, M. Voinov, and I. A. Grigor'ev, Free Radical Res. **45**, 417 (2011).
[44] K. Berg, M. Ericsson, M. Lindgren, and H. Gustafsson, PLoS One **9**, e90964 (2014).
[45] J. P. Gotham, R. Li, T. E. Tipple, Lancaster, Jr., T. M. Liu, and Q. Li, Free Radical Biol. Med. **154**, 84 (2020).

[46] P. Danhier, G. De Preter, S. Boutry, I. Mahieu, P. Leveque, J. Magat, V. Haufroid, P. Sonveaux, C. Bouzin, O. Feron *et al.*, Contrast Media Mol. I. **7**, 302 (2012).

[47] P. Danhier and B. Gallez, Contrast Media Mol. I. **10**, 266 (2015).